\documentclass[11pt]{article}

\pdfoutput=1

\usepackage[margin=2cm]{geometry}
\usepackage[citestyle=numeric,sorting=none,backend=bibtex]{biblatex}
\usepackage{setspace}
\usepackage[utf8]{inputenc}
\usepackage[T1]{fontenc}
\usepackage{graphicx}
\usepackage{grffile}
\usepackage{longtable}
\usepackage{wrapfig}
\usepackage{rotating}
\usepackage[normalem]{ulem}
\usepackage{amsmath}
\usepackage{textcomp}
\usepackage{amssymb}
\usepackage{capt-of}
\usepackage{hyperref}
\usepackage{amsthm}
\usepackage{color}
\usepackage{qmech}
\usepackage{ylmath}

\usepackage{subcaption}
\usepackage{tikz}
\usetikzlibrary{shapes,
  arrows.meta,
  calc,
  fit,
  decorations.pathmorphing,
  decorations.markings,
  decorations.pathreplacing}

\usepackage{authblk}

\newcommand{\haarexp}{\mathbb{E}_H}
\newcommand{\HS}{\mathscr{H}}
\newcommand{\SymG}{\mathbb{S}}
\newcommand{\swap}{\mathbb{X}}

\hypersetup{
 pdfauthor={Yuri D. Lensky and Xiao-Liang Qi},
 pdftitle={Chaos and High Temperature Pure State Thermalization},
 pdfkeywords={},
 pdfsubject={},
 pdfcreator={}, 
 pdflang={English}}

\theoremstyle{plain}
\newtheorem{theorem}{Theorem}

\begin{document}

\title{Chaos and High Temperature Pure State Thermalization}
\author[1]{Yuri D. Lensky \thanks{\href{mailto:ydl@stanford.edu}{ydl@stanford.edu}}}
\author[1,2]{Xiao-Liang Qi}
\affil[1]{\small \em  Stanford Institute for Theoretical Physics, Stanford University, Stanford CA 94305 USA}
\affil[2]{\small \em  Institute for Advanced Study, Princeton NJ 08540 USA}

\maketitle

\begin{abstract}
  Classical arguments for thermalization of isolated systems do not apply in a straightforward way to the quantum case. Recently, there has been interest in diagnostics of quantum chaos in many-body systems. In the classical case, chaos is a popular explanation for the legitimacy of the methods of statistical physics. In this work, we relate a previously proposed criteria of quantum chaos in the unitary time evolution operator to the entanglement entropy growth for a far-from-equilibrium initial pure state. By mapping the unitary time evolution operator to a doubled state, chaos can be characterized by suppression of mutual information between subsystems of the past and that of the future. We show that when this mutual information is small, a typical unentangled initial state will evolve to a highly entangled final state. Our result provides a more concrete connection between quantum chaos and thermalization in many-body systems.

\end{abstract}

\section{Introduction}
\label{sec:intro}
Empirically, there is a generic tendency towards entropy growth in many body systems. This ``arrow of time'' appears at odds with the fact that our physical models of these systems are often time-reversal symmetric. A common argument has been available from the time of Boltzmann: states are exponentially likely to evolve to states of higher entropy, simply due to the counting of states at given entropy (see for example \autocite{landau1969statistical}). In fact, in the presence of time-reversal symmetry this explanation is incomplete. For every state of a given entropy with entropy growth, there is a state with the same entropy but with entropy decay. The best we can actually hope for is to explain why some class of preferred states experiences entropy growth in some subsystem under specific dynamics.

This question is closely related to another aspect of quantum many body systems. The use of statistical ensembles to understand the long-time collective behavior of many degrees of freedom in terms of local microscopic interactions is one of the great simplifications and triumphs of modern physics. It is natural to ask when and why the use of these ensembles is justified; if all expectation values of interest for a given initial condition after time evolution can be computed to arbitrary accuracy in an ensemble depending only on macroscopic parameters, we will say that system thermalizes for those initial conditions and that time evolution. For isolated classical systems, dynamical chaos is a sufficient and generic condition for ergodicity in phase space, which explains the accuracy of the microcanonical ensemble and hence equilibrium statistical mechanics.

The quantum case is more complicated. It is important to note that there are classical systems with few degrees of freedom whose observables are well-described by the microcanonical ensemble, but upon quantization do not thermalize. Although there are experimental probes of few-body quantum systems that \emph{do} thermalize \autocite{neill2016ergodic}, generic thermalization in quantum systems appears to be inherently a many-body effect \autocite{deutsch1991quantum}. Quantum mechanics also supports a long-time behavior not present in classical systems, Many-Body Localization (MBL). Recent experiments detecting these phases \autocite{Choi1547} provide practical motivation to explain the mechanism of and conditions for thermalization of isolated systems. Finally, our understanding of quantum chaos and ergodicity is still incomplete, and leading justifications for quantum statistical mechanics are not as directly connected to quantum chaos as in the classical case.

There are two leading explanations for quantum thermalization. One is known as Canonical Typicality (CT) \autocite{goldstein2006canonical,popescu2006entanglement}, the statement that due to the exponentially large dimension of Hilbert space or the subspaces associated to finite energy windows, almost all pure states in the subspace will appear as if they were randomly chosen from that subspace, i.e. indistinguishable from the microcanonical ensemble, on any small subregion. Importantly, the CT approach can be extended to the dynamical result \autocite{linden2009quantum} that, under weak assumptions about the distribution of eigenvalues of the Hamiltonian, a subsystem interacting with a sufficiently large bath will spend most of its time close to its time average, independent of the initial state of the subsystem and for almost all initial states of the bath. One useful way to view CT is as an extension of the statistical argument for entropy growth: most states in a subspace are already close to maximum entanglement within the subspace. The other explanation is the Eigenstate Thermalization Hypothesis (ETH) \autocite{deutsch1991quantum,srednicki1994chaos}, which loosely stated is the conjecture that the high-energy eigenstates of quantized classically chaotic systems are indistinguishable from the microcanonical ensemble of the same system for local observables. This conjecture is well-supported numerically for a large class of systems, and gives a very clean description of quantum thermalization when it applies.

There are some deficiencies remaining in both approaches. There are no direct criteria to evaluate on a Hamiltonian to see if ETH holds, short of finding the eigenstates. On a related note, ETH has only been proven true for a small class of systems. Finally, although the ETH is inspired by ideas about classical chaos and ergodicity, there is no proof that chaos in dynamics implies ETH.\footnote{Some interesting progress along this direction is the relation between ETH and out-of-time-order correlation functions in \autocite{huang17:finite}.} The conclusions of CT appear completely unrelated to whether a system is chaotic. The principle mechanism of CT is that typical states are close to maximally entangled, or already at equilibrium. The problem is that we would like information about highly atypical states (out-of-equilibrium low-entanglement states) that form a set of measure zero in most subspaces of high dimension. The dynamical extension \autocite{linden2009quantum} solves this problem for systems where a small subsystem can be highly atypical in a much larger typical bath. This is a reasonable assumption for a near-isolated quantum system interacting with the rest of the world, but is not useful when we wish to consider an even smaller class of states, where the entire system is far from equilibrium and has low entanglement. It is also a statement about time averages as opposed to instantaneous density matrices. Finally, like ETH, CT gives no criterion on the time evolution to distinguish thermalizing from non-thermalizing systems. On a related note, there is no explanation for the \emph{mechanism} of thermalization, apart from the high dimension of Hilbert space.

In this work, we link the entropy growth of low-entanglement states under unitary time \(U(t)\) to measures of quantum chaos associated with \(U(t)\). In doing so, we begin to address the above deficiencies in explanations of quantum thermalization. More specifically, we consider the chaos criteria proposed in \autocite{xlqi-bounds}. In this work, the unitary time evolution operator $U(t)=e^{-itH}$ is mapped to a doubled state. The doubled state is defined by considering two copies of the physical system and preparing maximally entangled EPR pairs between each site of the physical system and its doubling partner. Denoting this state by $\ket{I}$, the unitary $U(t)$ is mapped to the pure state $\matid \otimes U(t) \ket{I}$. By construction, the two copies of systems (named as the past system and future system) are always maximally entangled, with $U(t)$ the Schmidt matrix of the wavefunction. Quantum chaos is characterized by the suppression of mutual information between subsystems of the future and past systems. A small mutual information between a region $A$ in the past and a region $B$ in the future tells us that operators in $A$ mostly evolve to non-local operators exceeding the boundary of $B$, causing a suppression of local correlation functions. This criteria is shown to be related to another chaos criteria, the out-of-time-ordered correlation (OTOC) functions~\autocite{larkin1969quasiclassical,roberts2014two,Maldacena_2016,kitaevsyktalk,kitaevhiddenhawking}.

In this paper, we show that the mutual information criteria defined in \autocite{xlqi-bounds} also controls the entropy growth for proper choices of low entropy initial states. More specifically, we consider a given partition of the system into multiple regions, and consider an ensemble of initial states that are tensor products of random states in each region. After time evolution by $U(t)$, we study the purity $\Tr \rho_A(t)^2$ of a subsystem $A$ in the final state. The ensemble average of the second Renyi entropy is determined by a sum in which each term is controlled by the second Renyi mutual information in the doubled state. When the mutual information terms are sufficiently small, a typical product state at initial time evolves into a state with nearly maximal entropy. Therefore we have shown that chaos in the dynamics $U(t)$ implies thermalization, at least for the ensemble of unentangled initial states we define. Since the random product state has a high energy, the final state has maximal entropy and infinite temperature. We also discuss the generalization of our result to initial state ensembles with finite temperature.

The remainder of the paper is organized as follows. In Section~\ref{sec:quantum_chaos}, we review the relevant aspects of quantum chaos and show how information theoretic quantities are linked to re-thermalization of the thermal ensemble. Sections~\ref{sec:random-product} and \ref{sec:restricted-products} are the main results of this work, and demonstrate a connection between thermalization of product states, quantum information theory, and quantum chaos. Derivations of the main results are in Appendix~\ref{sec:deriv-main-results}.

In what follows, we denote density matrices of subsystems by \(\rho_A = \Tr_{\bar{A}} \rho\), dimension of subsystem \(A\) by \(D_A\), and operators that act by identity in \(\bar{A}\) as \(O_A\). As a reminder, the von Neumann entropy of a density matrix \(\rho\) is \(S[\rho] = - \Tr[\rho \ln \rho]\), and the mutual information between two subsystems in \(\rho\) is \(I[\rho; A, B] = S[\rho_A] + S[\rho_B] - S[\rho_{A \cup B}]\).
\section{Quantum chaos}
\label{sec:quantum_chaos}
We start by reviewing some recent results in understanding quantum chaos. In classical systems, one diagnostic of chaos is exponential sensitivity to initial conditions, quantified by the exponential growth of the Poisson bracket of some pair of phase space coordinates \(\{q(t), p(0)\} = \partial q(t) / \partial q(0) \sim e^{\lambda_L t}\). A natural generalization to the quantum case is \(\expv{-[O_A(t), O_B(0)]^2}_{\rho}\) \autocite{larkin1969quasiclassical}. Of the four terms in the expansion of this expression, the most interesting for our purpose is the out-of-time ordered four-point correlator (OTOC)
\begin{equation}
\label{eq:4pt-otoc-def}
C_4(O_A(t), O_B(0))_{\rho} = \expv{O_A(t) O_B(0) O_A(t) O_B(0)}_{\rho},
\end{equation}
whose decay for thermal \(\rho\) has been interpreted as a signal of quantum chaos \autocite{roberts2014two,Maldacena_2016,xlqi-bounds}.

\begin{figure}[htb]
  \centering
  \begin{subfigure}[b]{0.25\textwidth}
    \centering
    \begin{tikzpicture}
      \begin{scope}[scale=2]
        \draw[line width=3] (0, 1) -- (1, 1);
        \draw[thin] (1, 1) -- (1, 0) -- (0, 0) -- (0, 1);
        \foreach \x in {0,...,7} {
          \draw ($(\x/7, 1)$) -- ($(\x/7, 1.35)$);
          \draw ($(\x/7, 0)$) -- ($(\x/7, -0.35)$);
        }
        \node at (0.5, 0.5) {$U(t)$};
        \draw[thick,
        decoration={
          brace,
          mirror,
          amplitude=7
        },
        decorate,
        postaction={
          decoration=
          {markings,
            mark=at position 0.5 with { \node[yshift=-7] {$\HS^*$}; }},
          decorate}]
        (0, -0.50) -- (1, -0.50);
        \draw[thick,
        decoration={
          brace,
          amplitude=7
        },
        decorate,
        postaction={
          decoration=
          {markings,
            mark=at position 0.5 with { \node[yshift=7] {$\HS$}; }},
          decorate}]
        (0, 1.5) -- (1, 1.5);
      \end{scope}
    \end{tikzpicture}
    \caption{Time evolution $U(t)$.}
    \label{fig:u-as-operator}
  \end{subfigure}%
  \begin{subfigure}[b]{0.3\textwidth}
    \centering
    \begin{tikzpicture}
      \begin{scope}[scale=2]
        \draw[line width=3] (0, 1) -- (1, 1);
        \draw[thin] (1, 1) -- (1, 0) -- (0, 0) -- (0, 1);
        \foreach \x in {7,...,0} {
          \draw ($(\x/7, 1)$) -- ($(\x/7, 1.35)$);
          \draw ($(\x/7, 0)$) -- ($(\x/7, -0.15)$);

          \draw[preaction={draw, line width=3, white}] ($(\x/7, -0.15)$)
          -- ($(\x/7 - 0.6, -0.7)$)
          -- ($(\x/7 - 1.2, -0.15)$)
          -- ($(\x/7 - 1.2, 1.35)$);
        }
        \node at (0.5, 0.5) {$U(t)$};
        \draw[thick,
        decoration={
          brace,
          amplitude=7
        },
        decorate,
        postaction={
          decoration=
          {markings,
            mark=at position 0.5 with { \node[yshift=7] {``Future'' $\HS$}; }},
          decorate}]
        (0, 1.5) -- (1, 1.5);
        \draw[thick,
        decoration={
          brace,
          amplitude=7
        },
        decorate,
        postaction={
          decoration=
          {markings,
            mark=at position 0.5 with { \node[yshift=7] {``Past'' $\HS$}; }},
          decorate}]
        (-1.2, 1.5) -- (-0.2, 1.5);
        \path[fill opacity=0.1, fill=gray, draw opacity=0.2, draw=black] (-1.3, -1.1) rectangle (1.1, -0.075);
        \node at ($0.5*(1.1 - 1.3,-1.8)$) {$\ket{I}$};
      \end{scope}
    \end{tikzpicture}
    \caption{$U(t)$ as a state in $\HS \otimes \HS$.}
    \label{fig:u-as-state}
  \end{subfigure}%
  \begin{subfigure}[b]{0.4\textwidth}
    \centering
    \begin{tikzpicture}
      \node at (-2.1, -.8) {$\Tr_{\overline{A}_F \cup \overline{B}_P}$};
      \begin{scope}
        \begin{scope}
          \draw[line width=2] (0, 1) -- (1, 1);
          \draw[thin] (1, 1) -- (1, 0) -- (0, 0) -- (0, 1);
          \foreach \x in {7,...,0} {
            \draw ($(\x/7, 1)$) -- ($(\x/7, 1.35)$);
            \draw ($(\x/7, 0)$) -- ($(\x/7, -0.15)$);

            \draw[preaction={draw, line width=2, white}] ($(\x/7, -0.15)$)
            -- ($(\x/7 - 0.6, -0.7)$)
            -- ($(\x/7 - 1.2, -0.15)$)
            -- ($(\x/7 - 1.2, 1.35)$);
          }
          \node at (0.5, 0.5) {$U(t)$};
          \path[draw=black, fill=red, fill opacity=0.1] ($(2/7-1.27,1.2)$) rectangle ($(4/7-1.27, 1.45)$);
          \path[draw=black, fill=red, fill opacity=0.1] ($(5/7-1.27,1.2)$) rectangle ($(8/7-1.27, 1.45)$);
          \path[draw=black] ($(1/14,1.2)$) rectangle ($(9/14, 1.45)$);
        \end{scope}
        \begin{scope}[yshift=-1.5cm,y=-1cm]
          \draw[line width=2] (0, 1) -- (1, 1);
          \draw[thin] (1, 1) -- (1, 0) -- (0, 0) -- (0, 1);
          \foreach \x in {7,...,0} {
            \draw ($(\x/7, 1)$) -- ($(\x/7, 1.35)$);
            \draw ($(\x/7, 0)$) -- ($(\x/7, -0.15)$);

            \draw[preaction={draw, line width=2, white}] ($(\x/7, -0.15)$)
            -- ($(\x/7 - 0.6, -0.7)$)
            -- ($(\x/7 - 1.2, -0.15)$)
            -- ($(\x/7 - 1.2, 1.35)$);
          }
          \node at (0.5, 0.5) {$U(t)^{\dagger}$};
          \path[draw=black, fill=red, fill opacity=0.1] ($(2/7-1.27,1.2)$) rectangle ($(4/7-1.27, 1.45)$);
          \path[draw=black, fill=red, fill opacity=0.1] ($(5/7-1.27,1.2)$) rectangle ($(8/7-1.27, 1.45)$);
          \path[draw=black] ($(1/14,1.2)$) rectangle ($(9/14, 1.45)$);
        \end{scope}
      \end{scope}
      \node at (1.3, -.8) {$=$};
      \begin{scope}[xshift=2.5cm]
        \begin{scope}
          \draw[line width=3] (0, 1) -- (1, 1);
          \draw[thin] (1, 1) -- (1, 0) -- (0, 0) -- (0, 1);
          \foreach \x in {7,6,5,3,2} {
            \draw ($(\x/7, 0)$) -- ($(\x/7, -0.15)$);

            \draw[preaction={draw, line width=2, white}] ($(\x/7, -0.15)$)
            -- ($(\x/7 - 0.6, -0.7)$)
            -- ($(\x/7 - 1.2, -0.15)$)
            -- ($(\x/7 - 1.2, 1.35)$);
          }
          \foreach \x in {4,1,0} {
            \draw[preaction={draw, line width=2, white}] ($(\x/7, 0)$) -- ($(\x/7, -1)$);
          }
          \foreach \x in {1,...,4} {
            \draw ($(\x/7, 1)$) -- ($(\x/7, 1.35)$);
          }
          \node at (0.5, 0.5) {$U(t)$};
          \foreach \x in {7,...,5,0} {
            \draw ($(\x/7, 1)$) -- ++(0, .15) coordinate (a);
            \draw[preaction={draw, line width=2, white}]
            (a) -- ++(1/28, 0.1) -- ++(1/28, -0.1) -- ++(0, -3);
          }
        \end{scope}
        \begin{scope}[yshift=-1.5cm,y=-1cm]
          \draw[line width=3] (0, 1) -- (1, 1);
          \draw[thin] (1, 1) -- (1, 0) -- (0, 0) -- (0, 1);
          \foreach \x in {7,6,5,3,2} {
            \draw ($(\x/7, 0)$) -- ($(\x/7, -0.15)$);

            \draw[preaction={draw, line width=2, white}] ($(\x/7, -0.15)$)
            -- ($(\x/7 - 0.6, -0.7)$)
            -- ($(\x/7 - 1.2, -0.15)$)
            -- ($(\x/7 - 1.2, 1.35)$);
          }
          \foreach \x in {4,1,0} {
            \draw[preaction={draw, line width=2, white}] ($(\x/7, 0)$) -- ($(\x/7, -1)$);
          }
          \foreach \x in {1,...,4} {
            \draw ($(\x/7, 1)$) -- ($(\x/7, 1.35)$);
          }
          \node at (0.5, 0.5) {$U(t)^{\dagger}$};
          \foreach \x in {7,...,5,0} {
            \draw ($(\x/7, 1)$) -- ++(0, .15) coordinate (a);
            \draw[preaction={draw, line width=2, white}]
            (a) -- ++(1/28, 0.1) -- ++(1/28, -0.1) -- ++(0, -3);
          }
        \end{scope}
      \end{scope}
    \end{tikzpicture}
    \caption{Density matrix $\rho^{U(t)}_{A_F \cup B_P}$; on the left, $A$ has a white box around it, and $B$ has a red box.}
    \label{fig:rho-u}
  \end{subfigure}
  \caption{Pictorial representation and explicit construction of the mapping from time evolution operator $U(t)$ to the state $\ket{U(t)}\in \HS \otimes \HS$ and the associated density matrix $\rho^U(t)$. First, in \ref{fig:u-as-operator} we introduce our notation and draw $U(t)$ as a tensor with ``input'' legs at the bottom and ``output'' legs at the top. To help keep track of the future and past, we draw the output edge of $U(t)$ as a bolded line. Each leg corresponds to a subsystem of $\HS$ and denotes an index in the tensor, and contraction is represented by simply connecting ``input'' with ``output'' legs. A particular example of this operation is shown in \ref{fig:u-as-state}, where we depict action by $\matid \otimes U(t)$ on the maximally entangled state $\ket{I}$, turning $U(t)$ into a state $\ket{U(t)}$ on a doubled system. In \ref{fig:rho-u} we show $\rho^{U(t)}_{A_F \cup B_P} = \Tr_{\overline{A}_F \cup \overline{B}_P} \ket{U(t)}\bra{U(t)}$. It is clear from this construction that $\rho^{U(t)}$ is maximally entangled between the past and future, so that for any region $R$ exclusively in the past or future, $S[\rho^{U(t)}_{R}] = S^{(2)}[\rho^{U(t)}_{R}] = \ln D_R$.}
  \label{fig:explicit-rho-u-construction}
\end{figure}
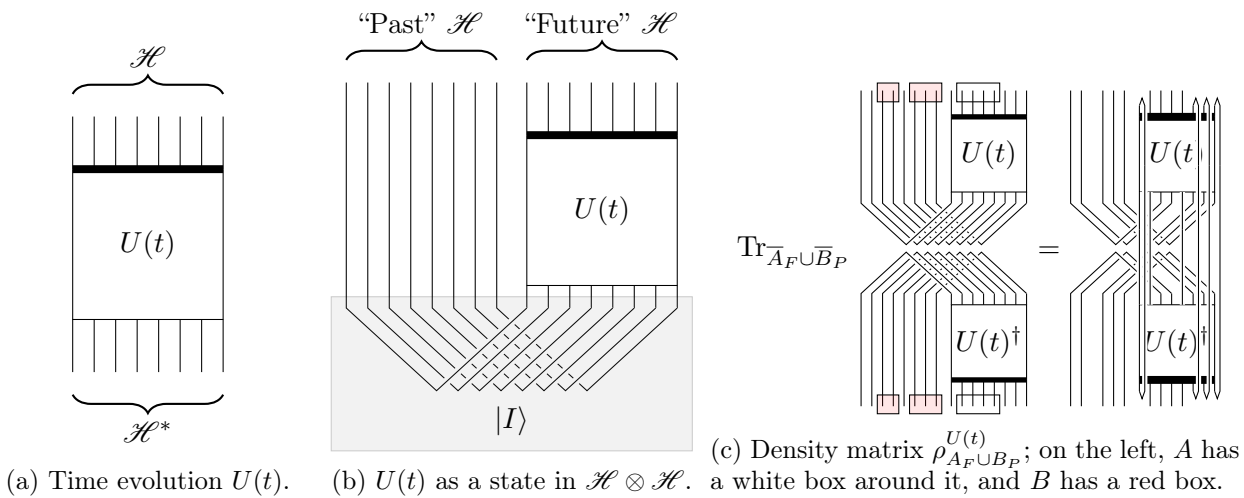

The decay of \(C_4\) seems to be an operator-dependent statement, but is in fact related to an information-theoretic quantity, the second Renyi entropy \(S^{(2)}\), computed from the time evolution \(U(t) = e^{- i H t}\) \autocite{xlqi-bounds}. Since \(U(t) \in \HS \otimes \HS^*\), we can consider it as a normalized state in a Hilbert space with inner product \(\langle A, B \rangle = \Tr[A^{\dagger} B] / D\) (here $D$ is the dimension of $\HS$). We can associate the copy of the Hilbert space corresponding to the future (past) with the left (right) tensor factor in \(\HS \otimes \HS^*\). For intuition and computation, it can be useful to choose an isomorphism $\HS^* \simeq \HS$ that is compatible with the tensor factorization and think of $U(t)$ as an entangled state on two copies of the original system. Concretely, we consider the original Hilbert space $\HS$ as a tensor product of small Hilbert spaces (for example on each site of a lattice): $\HS=\bigotimes_x \HS_x$. Denoting an orthonormal basis of $\HS_x$ by $\ket{\alpha_x}, \alpha_x=1,2,...,\dim \HS_x$, one can define the maximally entangled state in the doubled Hilbert space as $\ket{I}=\bigotimes_x \left(D^{-1/2}\sum_{\alpha_x = \alpha_x^P = \alpha_x^F}\ket{\alpha_x^P}\otimes \ket{\alpha_x^F}\right)$. The state $\ket{I}$ encodes an isomorphism from operators (elements of $\HS^* \otimes \HS$) to states in a doubled system (elements of $\HS \otimes \HS$) by right action, so we can explicitly map the unitary operator $U(t)$ to the state $\ket{U(t)} = (\matid \otimes U(t)) \ket{I}$.\footnote{It is important to note that $\ket{U(t)}$ depends on the basis choice used when defining $\ket{I}$, which determines the isomorphism $\HS \simeq \HS^*$.
  However, entanglement properties of $\rho^{U(t)}$ are completely independent from the basis choice.
} We denote the density matrix associated with the pure state $\ket{U(t)}$ as $\rho^{U(t)}=\ket{U(t)}\bra{U(t)}$. The construction of $\ket{U(t)}$ and an example partial trace of $\rho^{U(t)}$ is illustrated in Figure~\ref{fig:explicit-rho-u-construction}.

Correlations between the past and future copies of Hilbert space in \(U(t)\) are related to chaos and scrambling. For example, the mutual information between a region \(A\) in the future and region \(B\) in the past bounds correlations in time:
\begin{equation}
  \label{eq:i-rhou-bound}
  I[\rho^{U(t)}; A_F, B_P] \ge \frac{1}{2} \left( \frac{\expv{O_A(t) O_B(0)}_{\beta = 0} - \expv{O_A(t)}_{\beta = 0} \expv{O_B(0)}_{\beta = 0}}{\| O_A \| \|O_B \|}\right)^2.
\end{equation}
We can already see a connection of the information content of \(\rho^{U(t)}\) and thermalization in \eqref{eq:i-rhou-bound}. If the mutual information between \(A\) in the future and \(B\) in the past is small in \(\rho^{U(t)}\), the action of an operator in \(B\) in the past has no influence on the action of an operator in \(A\) in the future. This shows that small mutual information in \(\rho^{U(t)}\) is sufficient for re-thermalization of the infinite temperature ensemble after perturbation. Thus in this case, we have the natural statement that information in $U(t)$ between regions in the future and past tells us how sensitive the future region is to the initial conditions in the past region. The main goal of this work is to extend this result to far-from-equilibrium pure states. More generally, \eqref{eq:i-rhou-bound} shows that we can think of the mutual information \(I[\rho^{U(t)}; A_F, B_P]\) roughly as quantifying how much initial conditions in \(B\) determine the subsystem \(A\) after time evolution \(U(t)\).

A main result of \autocite{xlqi-bounds} is a more explicit connection of past-future mutual information to chaos: the average of the OTOC (equation \eqref{eq:4pt-otoc-def}) over operators in subsystem \(A\) in the future and \(B\) in the past is proportional to \(S^{(2)}[\rho^{U(t)}_{A_F \cup \bar{B}_P}]\), where $\bar{B}_P$ is the complement of $B_P$ in the past system. By average of operators on a subsystem \(A\) we mean a weighted sum over the \(D_A^2\) Hermitian operators in a complete, orthonormal basis (under the above inner product on operators). We will write these operators in a script font, as \(\mathcal{O}_A\), with the average implied wherever they appear. The second Renyi entropy is defined as
\begin{equation}
  \label{eq:2nd-renyi-entropy}
  S^{(2)}[\rho] = - \log \Tr \rho^2
\end{equation}
and is a measure of uncertainty in \(\rho\): for pure states, \(S^{(2)} = 0\), while for maximally mixed states \(S^{(2)} = S = \log D\) where \(S\) is the von Neumann entropy and \(D\) is the dimension of the Hilbert space. There are some other properties of \(S^{(2)}[\rho^{U(t)}_{A_F \cup B_P}]\) that will be important in what follows. It can be seen from Jensen's inequality that \(S^{(2)}[\rho] \le S[\rho]\). Thus when \(S^{(2)}\) is near-maximal, so is \(S\). We also have the bounds \footnote{As can be seen by writing \(e^{-S^{(2)}[\rho^{U(t)}_{A_F \cup B_P}]} = \Tr[\pi_A(t) \pi_B(0)] / D_A D_B\) where \(\pi_A\) is the projector onto Hermitian operators in \(A\) in the operator Hilbert space defined above, or by the positivity of \(I^{(2)}\).} \(\ln D_A + \ln D_B \ge S^{(2)}[\rho^{U(t)}_{A_F \cup B_P}] \ge | \ln D_A - \ln D_B |\). Thus for \(B\) much larger than \(A\), \(S^{(2)}\) is large for ``kinematic'' reasons, independent of the time evolution. The more interesting quantity in this case is a version of the mutual information adopted for Renyi entropy, \(I^{(2)}[\rho^{U(t)}; A_F, B_P] = S^{(2)}[\rho^{U(t)}_{A_F}] + S^{(2)}[\rho^{U(t)}_{B_P}] - S^{(2)}[\rho^{U(t)}_{A_F \cup B_P}]\), which is non-negative in our state since \(\rho^{U(t)}\) is maximally entangled between the future and past. \(I^{(2)}[\rho^{U(t)}; A_F, B_P]\) captures fluctuations of \(S^{(2)}[\rho^{U(t)}_{A_F \cup B_P}]\) about its kinematic value, and bounds the corresponding mutual information \(I[\rho^{U(t)}; A_F, B_P]\) from above.

We can then write (as shown in \autocite{xlqi-bounds})
\begin{equation}
  \label{eq:unitary-renyi-otoc}
  D_A^2 C_4(\mathcal{O}_A(t), \mathcal{O}_{\bar{B}}(0))_{\beta = 0} = \exp(I^{(2)}[\rho^{U(t)}; A_F, B_P]).
\end{equation}
Thus scrambling in \(U(t)\) as quantified by \(S^{(2)}\) is directly related to chaos. A generically small four-point correlator means a large Renyi entropy or a small \(I^{(2)}\). For \(A\) and \(B\) small, the expression \eqref{eq:unitary-renyi-otoc} is actually in terms of non-local operators on \(\bar{B}\). In this case, there is a more natural expression in terms of local operators,
\begin{equation}
  \label{eq:unitary-renyi-2pt}
  (D_A D_B \expv{\mathcal{O}_A(t) \mathcal{O}_B(0)}_{\beta = 0})^2 = \exp(I^{(2)}[\rho^{U(t)}; A_F, B_P]).
\end{equation}
The two expressions \eqref{eq:unitary-renyi-otoc} and \eqref{eq:unitary-renyi-2pt} emphasize the important point that the second Renyi mutual information characterizes the behavior of both two-point functions and OTOC. $I^{(2)}[\rho^{U(t)}; A_F, B_P]$ between two small regions $A$ and $B$ is governed by two-point functions of operators supported on $A$ and $B$, while that between a small region $A$ and a big region $B$ (bigger than half system size) is governed by the OTOC of operators supported on $A$ and the smaller region $\bar{B}$.
This is also consistent with the fact that the decay of the OTOC implies a stronger scrambling of information than simply the decay of two-point functions. As we will see below, one utility of the point of view of information is a unified treatment of the two- and four-point functions.

In \autocite{xlqi-bounds}, the relationship \eqref{eq:unitary-renyi-otoc} is used to show that a four-point correlator decaying to some value less than \(\epsilon\) in any region implies that the sum of mutual informations \(I[\rho^{U(t)}; A_F, B_P] + I[\rho^{U(t)}; A_F, \bar{B}_P] \le I^{(2)}[\rho^{U(t)}; A_F, B_P] + I^{(2)}[\rho^{U(t)}; A_F, \bar{B}_P] < 4 \ln D_A + 2 \ln \epsilon\). In principle, \(\epsilon\) can be so small that this sum is arbitrarily close to zero. In more realistic models, we can expect that the OTOC will decay as some polynomial of the logarithm of the \emph{total} Hilbert space dimension. This sum \(I[\rho^{U(t)}; A_F, B_P] + I[\rho^{U(t)}; A_F, \bar{B}_P]\) minus $I[\rho^{U(t)}; A_F,B_P\bar{B}_P]\equiv2 \ln D_A$ is called tripartite information and its negativity is proposed as a measure of ``scrambling'' due to unitary time evolution; then quantum chaos as measured by the decay of \(C_4\) implies scrambling.

We would like to make a side remark at the end of this section. We treat \(I^{(2)}\) and \(I\) for \(\rho^{U(t)}\) as operator-independent diagnostics of chaos. It is clear from the discussion above that if the OTOC and two-point functions decay generically, \(I^{(2)}\) will be small, which implies \(I\) is small as well. Although it is most direct from the discussion above to treat \(I^{(2)}\) as the intrinsic measure of chaos and \(I\) simply as a quantity \emph{also} small in chaotic systems only because it is bounded by \(I^{(2)}\), the true mutual information \(I\) is more natural in many other contexts and it is intuitive that small mutual information of \(\rho^{U(t)}\) should \emph{imply} chaos. To that end, using a bound on von Neumann entropy in terms of Renyi entropy \autocite{harremoes2001inequalities} (see Appendix~\ref{sec:deriv-mi-bound}), we can show that
\begin{equation}
  \label{eq:i-bound-i2}
  \exp(I^{(2)}[\rho^{U(t)}; A_F, B_P]) \le 1 + \left( 1 + \frac{1}{\ln (D_A D_B)} \right) (D_A D_B - 1) \frac{I[\rho^{U(t)}; A_F, B_P]}{\ln (D_A D_B)}.
\end{equation}
Thus a sufficiently small mutual information implies small \(I^{(2)}\), which in turn implies chaos according to the OTOC. In the remainder of the work, we will focus on \(I^{(2)}\), but \eqref{eq:i-bound-i2} should be kept in mind as a way to bound \(I^{(2)}\) in terms of the true mutual information.

\section{Thermalization of completely random product states}
\label{sec:random-product}
Our goal is to understand how entropy growth and thermalization is related to quantum chaos as defined above. As discussed in Section \ref{sec:intro}, entropy growth is a state-dependent statement and can only be true for specific classes of states, for example initial states with small entanglement. The most naive choice of initial state ensemble is product states of some fixed granularity. More precisely, we consider a partition of the initial system into regions $R_s$ such that $\cup_{s=1}^{S}R_s=P$ is the whole system. Correspondingly, each region $R_s$ has a Hilbert space $\HS_s$, and the Hilbert space of the whole system can be written as a tensor product of subsystems \(\HS = \bigotimes_s \HS_s\). We consider states of the form \(\ket{\psi(0)} = \bigotimes_s \ket{a_s}\), with \(\ket{a_s}\) a random pure state in $\HS_s$. An example of one of these states, along with its time evolution, is shown in Figure~\ref{fig:random-pure-state}. There is no change in the global entropy of a density matrix under unitary evolution, but there \emph{can} be changes in subsystems. Thus we consider the second Renyi entropy of the density matrix corresponding to an initial low-entanglement pure state in some subsystem \(A\) \footnote{Note that \(A\) need not factorize through the \(\{\HS_s\}\).} after time evolution by \(U(t)\), averaged uniformly (according to the Haar measure) over initial states of the form \(\ket{\psi(0)}\). Denoting the average \(\haarexp[f(\bigotimes_s \ket{a_s})] = \int \prod_s dU_s f(\bigotimes_s U_s \ket{a_s})\), where integrals are done over the Haar measure, and \(\rho^{\psi}(0) = \ket{\psi(0)}\bra{\psi(0)}\), we find our main result
\begin{equation}
  \label{eq:infinite-temp-renyi2}
  \haarexp\left[\exp(- S^{(2)} [\rho^{\psi}_A(t)])\right]
  = \frac{1}{\prod_s (1 + 1 / D_s)} \frac{1}{D_A} 
  \left( 1 + \sum_{\substack{R \in \mathcal{P}(\{R_s\}) \\ R \ne \varnothing}} \frac{\exp(I^{(2)}[\rho^{U(t)}; A_F, R_P])}{D_R} \right)
\end{equation}
where the sum runs over all nontrivial subregions $R=R_{i_1}\cup R_{i_2}\cup \cdots \cup R_{i_n}$ that are unions of some of the building blocks $R_s$. \(\mathcal{P}(\{R_s\})\) denotes the set of all such $R$'s, {\it i.e.} the powerset of $\{R_1,R_2,...,R_S\}$. A similar relation has recently been studied in the context of random dynamics in~\autocite{You:2018ons}. A representation of a typical term in the sum is shown in Figure~\ref{fig:pow-set-term}. We give some examples of this formula below, and present a derivation in Appendix~\ref{sec:deriv-main-results}.

\tikzset{region-bracket/.style={thick,
    decoration={brace,
      amplitude=3,
    },
    decorate,
  }}

\tikzset{region-mirror-bracket/.style={thick,
    decoration={brace,
      amplitude=3,
      mirror,
    },
    decorate,
  }}

\begin{figure}[htb]
  \centering
  \begin{subfigure}[t]{0.3\textwidth}
    \centering
    \begin{tikzpicture}
      \begin{scope}[scale=2]
        \draw[line width=3] (0,1) -- (1,1);
        \draw[thin] (1,1) -- (1,0) -- (0,0) -- (0,1);
        \node at (0.5,0.5) {$U(t)$};

        \foreach \x in {0,...,7} {
          \draw ($(\x/7,1)$) -- ($(\x/7, 1.35)$)
          coordinate (output-leg-\x);
          \draw[-{Triangle[open,length=7,width'=0 1]}] ($(\x/7,0)$) -- ($(\x/7, -0.35)$)
          coordinate (input-leg-\x);
        }

        \draw[region-bracket] ([yshift=1]output-leg-2) -- ([yshift=1]output-leg-4)
        node[pos=0.5,yshift=10] {$A$};

        \path[fill opacity=0.1, fill=gray, draw opacity=0.2, draw=black]
        ([shift={(-0.1cm,0.2cm)}]input-leg-0) rectangle
        ([shift={(0.1cm,-0.3cm)}]input-leg-7)
        node[pos=0.5, fill opacity=1, yshift=-6] {$\ket{\psi(0)}$};
      \end{scope}
      \begin{scope}[xshift=-30pt,scale=2]
        \node at (0, 0.5) {%
          $
          \begin{tikzpicture}[baseline={([yshift=-.5ex]current bounding box.center)}]
            \draw[-{Triangle[open,length=7,width'=0 1]}] (0, 0.5) -- (0, 0);
          \end{tikzpicture}
          = \ket{a_s}$};
      \end{scope}
    \end{tikzpicture}
    \caption{Pictorial representation of the final state $\ket{\psi(t)}$ obtained from time evolution of the initial product state $\ket{\psi(0)}$.}
    \label{fig:random-pure-state}
  \end{subfigure}
  \begin{subfigure}[t]{0.3\textwidth}
    \centering
    \begin{tikzpicture}[scale=2]
        \draw[line width=3] (0,1) -- (1,1);
        \draw[thin] (1,1) -- (1,0) -- (0,0) -- (0,1);
        \node at (0.5,0.5) {$U(t)$};

        \foreach \x in {0,...,7} {
          \draw ($(\x/7,1)$) -- ($(\x/7, 1.35)$) coordinate (output-leg-\x);
          \draw ($(\x/7,0)$) -- ($(\x/7, -0.35)$) coordinate (input-leg-\x);
        }

        \draw[region-bracket] ([yshift=1]output-leg-2) -- ([yshift=1]output-leg-4)
        node[pos=0.5,yshift=10] {$A$};

        \foreach \firstleg / \lastleg [count=\irect from 1] in {2/3, 5/5, 7/7} {
          \node[inner sep=0.1cm, draw=black,fit={(input-leg-\firstleg) (input-leg-\lastleg)}] (r-rect-\irect) {};
        }

        \node at (0.5, -0.7) (r-label) {$R$};

        \foreach \x in {1,2,3} {
          \draw[-Stealth] (r-label) -- (r-rect-\x);
        }
    \end{tikzpicture}
    \caption{An example of region $R$ involved in the sum of Eq. \eqref{eq:infinite-temp-renyi2}.}
    \label{fig:pow-set-term}
  \end{subfigure}
  \begin{subfigure}[t]{0.3\textwidth}
    \centering
    \begin{tikzpicture}
      \begin{scope}[scale=2]
        \draw[line width=3] (0,1) -- (1,1);
        \draw[thin] (1,1) -- (1,0) -- (0,0) -- (0,1);
        \foreach \x in {0,...,7} {
          \draw ($(\x/7,1)$) -- ($(\x/7, 1.35)$) coordinate (output-leg-\x);
          \draw ($(\x/7,0)$) -- ($(\x/7, -0.35)$) coordinate (input-leg-\x);
        }
        \node at (0.5,0.5) {$U(t)$};

        \draw[region-bracket] ([yshift=1]output-leg-2) -- ([yshift=1]output-leg-4)
        node[pos=0.5,yshift=10] {$A$};

        \foreach \fl / \ll / \lbl in {0/2/s,3/7/sp} {
          \draw ([xshift=-0.04cm] input-leg-\fl) coordinate (\lbl-tri-tl)
          -- ([xshift=0.04cm] input-leg-\ll) coordinate (\lbl-tri-tr)
          -- ([yshift=-0.25cm] barycentric cs:\lbl-tri-tl=1,\lbl-tri-tr=1) coordinate (\lbl-tri-bot)
          -- cycle;
        }

        \draw[region-mirror-bracket] ([yshift=-0.27cm]s-tri-tl) -- ([yshift=-0.27cm]s-tri-tr)
        node[pos=0.5, yshift=-10] {$S$};

        \draw[region-mirror-bracket] ([yshift=-0.27cm]sp-tri-tl) -- ([yshift=-0.27cm]sp-tri-tr)
        node[pos=0.5, yshift=-10] {$\overline{S}$};
      \end{scope}
    \end{tikzpicture}
    \caption{The two-interval case considered in Eq. \eqref{eq:random-hight-bath-renyi2} and Eq. \eqref{eq:random-hight-bath-corr-fun-r2}.}
    \label{fig:2-region-state}
  \end{subfigure}
  \caption[Depiction of our setup.]{Depiction of our setup. In Figure~\ref{fig:random-pure-state} we show an example of a state from the ensemble we find the average entropy of in \eqref{eq:infinite-temp-renyi2}. Note in general each state $\ket{a_s}$ is different despite being drawn using the same symbol \tikz{\draw[-{Triangle[open,length=0 10, width'=0 1]}] (0, 0.3) -- (0, 0); }. In Figure~\ref{fig:pow-set-term}, we show a typical region $R$ from the powerset $\mathcal{P}(\{R_s\})$. Figure~\ref{fig:2-region-state} illustrates the special case with bipartition of the past system.}
  \label{fig:setup-formula}
\end{figure}
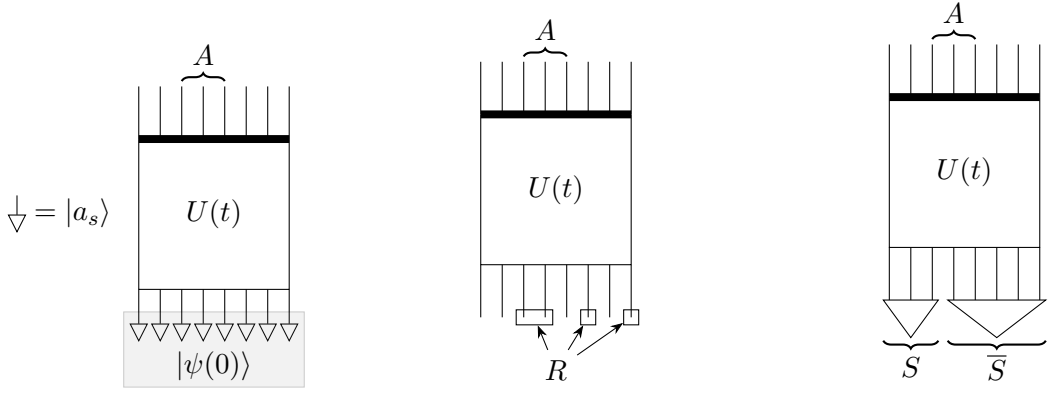

Apart from bounding von Neumann entropy from below, the utility of computing \(S^{(2)}\) is that it can be used to bound the one-norm difference of density matrices. Recall that \(|\expv{O}_{\rho} - \expv{O}_{\rho'}| / \|O\| \le \|\rho - \rho'\|_1\), so the one-norm is the natural distance for density matrices. By Jensen's inequality, 
\begin{equation}
  \label{eq:one-norm-inft-t}
  \haarexp \left[ \|\rho^{\psi}_A(t) - \rho_{A \beta = 0}\|_1 \right] \le \sqrt{D_A \haarexp e^{- S^{(2)}[\rho^{\psi}_A(t)]} - 1}.
\end{equation}
Thus as long as the deviation from maximal entanglement is sufficiently small, we can say a density matrix thermalizes in the one-norm in expectation. As we will see below, it turns out that thermalizing in expectation (at infinite temperature) sufficiently well is sufficient for most states to thermalize. Note that the infinite temperature ensemble is the appropriate choice here, since for typical Hamiltonians most states will be infinite temperature states (c.f. \autocite{Keating_2015}), and we always have \(\haarexp \expv{H}_{\rho^{\psi}} = \expv{H}_{\beta = 0}\).

To get an intuition for the implications of \eqref{eq:infinite-temp-renyi2}, it is illuminating to consider some special cases. First, we consider a trivial partition with only one region $R_1 = P$ equal to the whole system. In this case the ensemble is that of random pure states on the whole system. Our formula reduces to 
\begin{equation}
  \label{eq:random-pure-renyi2}
  \haarexp\left[\exp(- S^{(2)} [\rho^{\psi}_A(t)])\right] =
  \frac{1}{1 + 1/D} \frac{1}{D_A} \left( 1 + \frac{D_A^2}{D} \right),
\end{equation}
which is completely independent of dynamics. In the limit of large system size, as long as the subsystem \(A\) is less than half the system and grows at most linearly with system size, \(D_A^2 / D\) decays exponentially with system size. Then \eqref{eq:random-pure-renyi2} is the familiar statement that to exponential accuracy, a random pure state is close to maximally entangled in any small subsystem. This result is expected, as the typical pure state is indeed close to maximally entangled in any subsystem \autocite{lubkin1978entropy,PhysRevLett.72.1148,PhysRevLett.71.1291}, and a random state evolves to another random state under \emph{any} dynamics. In fact, the result of \autocite{lubkin1978entropy}, derived by explicit integration on \(S^{2D - 1}\), is a special case of \eqref{eq:random-pure-renyi2} for trivial evolution \(U(t) = \matid\). We can see the relationship to a more traditional measure of entanglement, the von Neumann entropy, by Jensen's inequality:
\begin{equation}
  \label{eq:random-pure-vonNeumann}
  \haarexp[S[\rho^{\psi}_A(t)]] \ge \haarexp[S^{(2)}[\rho^{\psi}_A(t)]]
  \ge \ln D_A - \left( \ln (1 + D_A^2 / D) - \ln (1 + 1/D) \right).
\end{equation}

We move to the case of two initial subsystems, \(\HS = \HS_S \otimes \HS_{\bar{S}}\) where we take \(1\ll D_S \le D_{\bar{S}}\). A typical state and its time evolution in this setup is shown in Figure~\ref{fig:2-region-state}. The expression \eqref{eq:infinite-temp-renyi2} becomes
\begin{multline}
  \label{eq:random-hight-bath-renyi2}
  - \ln \haarexp\left[\exp(- S^{(2)} [\rho^{\psi}_A(t)])\right]
  = \ln D_A 
  - \ln \left( 1 + \frac{e^{I^{(2)}[\rho^{U(t)}; A_F, S_P]}}{D_S} + \frac{e^{I^{(2)}[\rho^{U(t)}; A_F, \bar{S}_P]}}{D_{\bar{S}}} + \frac{D_A^2}{D} \right) \\
  + \ln (1 + (1 + D_S + D_{\bar{S}}) / D).
\end{multline}
Already in this next-to-simplest case dynamics play a central role. First, if region $S$ is large (and $\overline{S}$ is even larger), all the terms in \((1 + D_S + D_{\bar{S}}) / D\) are exponentially small. Regardless, this contribution serves to \emph{increase} \(S^{(2)}\). For \(A\) smaller than half system size, \(D_A^2 / D\) is exponentially small. The only decrease from maximal entanglement that can survive in the large system size limit is then due to the terms involving \(I^{(2)}[\rho^{U(t)}; A_F, R_P]\). Thus small \(I^{(2)}\) between \(A\) and both \(S\) and \(\bar{S}\), equivalent respectively to the generic decay of two- and four-point correlators between \(A\) and \(S\), is necessary and sufficient for the expectation of \(S^{(2)}[\rho^{\psi}_A(t)]\) to be near the maximal (equivalently thermal at infinite temperature) value for initial product states in \(S\) and \(\bar{S}\). It is important to note that ``small'' \(I^{(2)}\) depends on our choice of \(S\), as terms have the form \(e^{I^{(2)}[\rho^{U(t)}; A_F, R_P]}/ D_R\). If we want this contribution less than \(\epsilon_R\), we only require the condition that \(I^{(2)}[\rho^{U(t)}; A_F, R_P] < \ln (D_R \epsilon_R)\).

It is also useful to rewrite \eqref{eq:random-hight-bath-renyi2} in terms of correlation functions
\begin{multline}
  \label{eq:random-hight-bath-corr-fun-r2}
  - \ln \haarexp\left[\exp(- S^{(2)} [\rho^{\psi}_A(t)])\right]
  = \ln D_A \\
  - \ln \left( 1 +
    D_A^2 \left(\frac{1}{D} + D_S \langle \mathcal{O}_A(t) \mathcal{O}_S(0) \rangle_{\beta = 0} + \frac{C_4(\mathcal{O}_A(t), \mathcal{O}_S(0))_{\beta = 0}}{D_{\overline{S}}}\right) \right) \\
  + \ln (1 + (1 + D_S + D_{\bar{S}}) / D).
\end{multline}
For large systems $D_A^2 / D$ is exponentially smaller than $D_A^2 / D_{\overline{S}}$, so we can safely focus on the contributions due to correlators. It is clear that if the two-point functions decay and $D_A^2 / D_{\overline{S}}$ is finite, the deviation from maximum entanglement will be dictated by the (strictly positive) four-point term $D_A^2 C_4(\mathcal{O}_A(t), \mathcal{O}_S(0))_{\beta = 0} / D_{\overline{S}}$. Note that depending on the choice of $A$ and $S$, even with both less than half system size and (for lattice models) $|A| < |S|$, $D_A^2 / D_{\overline{S}}$ may be made of order, or even much greater than 1. As mentioned in Section~\ref{sec:quantum_chaos}, $C_4(\mathcal{O}_A(t), \mathcal{O}_S(0))_{\beta = 0}$ can be as small as an inverse polynomial in the logarithm of system dimension in chaotic systems, so as long as $A$ and $S$ are chosen so that $D_A^2 / D_{\overline{S}}$ does not grow too quickly with system size, in chaotic systems random product states on $S$ and $\overline{S}$ will evolve to look thermal in $A$. In the limit \(D_A^2 / D_{\bar{S}} \to 0\), only the local two-point function will contribute to deviations from \(\ln D_A\). This is the case considered by CT, and shows that chaos in the OTOC sense is not necessary for thermalization into a much larger random bath. On the other hand, when \(D_{\bar{S}}\) is finite, if four-point correlations do not decay sufficiently we can have significant corrections to thermal entropy.

This argument extends without significant modification to the case of \(S\) initial subsystems \(\HS = \bigotimes_{s = 1}^S \HS_s\), where \(S\) may grow linearly with system size. As long as two-point functions generically decay between \(A\) and subsystems up to half system size, the contribution from summands in \eqref{eq:infinite-temp-renyi2} involving \(R\) less than half the system will be small. The decay of four-point correlators between \(A\) and subsystems up to half system size is necessary to bound contributions from summands involving \(R\) greater than half system size. Concretely, we need, for regions $R$ less than half system size, $\expv{\mathcal{O}_A(t) \mathcal{O}_R(0)}_{\beta = 0} \ll 1 / (D_A \sqrt{D_R})$ and $C_4(\mathcal{O}_A(t) \mathcal{O}_R(0))_{\beta = 0} \ll D_{\overline{R}} / D_A^2$ for $A$ to look maximally entangled on average. Note that in an integrable system, it is expected that \(I^{(2)}\) will always be high for \emph{some} subextensive \footnote{By subextensive we do \emph{not} mean not growing with system size, but we mean of dimension \(D_R\) small enough that \(D_A^2 / D_R\), with \(A\) chosen as in the examples above, is not exponentially small.} subregions \(R\) (for example numerics in \autocite{xlqi-bounds}), although these regions may change in time as information propagates. Some subextensive region of initial conditions largely determines the density matrix in \(A\). This demonstrates an obstacle to thermalization in non-chaotic isolated systems. In contrast, a chaotic system will scramble information about the initial conditions in each \(\HS_s\) across extensive regions of the system. Equivalently, extensive knowledge of initial conditions determines the density matrix in \(A\). The only \(R\) for which \(\exp(I^{(2)}[\rho^{U(t)}; A_F, R_P]) \gtrsim D_A\) have \(D_R \sim D\), so that even if \(I^{(2)}[\rho^{U(t)}; A_F, R_P]\) attains its maximum \(2 \ln D_A\), \(D_A^2 / D_R\) decays exponentially in system size if \(A\) is chosen as in the examples above.

For systems with
\begin{equation}
  \label{eq:fluct-bound-epsilon}
  \epsilon = \haarexp\left[D_A e^{- S^{(2)}[\rho^{\psi}_A(t)]} - 1 \right] 
  =\frac{1}{\prod_s (1 + 1/D_s)} \sum_{\substack{R \in \mathcal{P}(\{R_s\}) \\ R \ne \varnothing}} \frac{e^{I^{(2)}[\rho^{U(t)}; A_F, R_P]}}{D_R}
  + \left( \frac{1}{\prod_s (1 + 1/D_s)} - 1 \right)
\end{equation}
small, we can meaningfully bound the number of states that do not thermalize by Markov's inequality:
\begin{equation}
  \label{eq:fluct-bound}
  \Pr \left( D_A e^{- S^{(2)}[\rho^{\psi}_A(t)]} - 1 > \delta \right) < \epsilon / \delta.
\end{equation}
The conclusion is that if we find that states are expected to thermalize sufficiently well, then a particular state is likely to have the average behavior after long times. This bound is easily ``weakened'' to a statement about probabilities of significant deviation for local entropy. On the other hand, if states are not expected to thermalize, we do not expect to find such a bound on physical grounds; the long-time trajectory of non-thermalizing systems can depend sensitively on the details of initial conditions.
\section{Finite temperature extension}
\label{sec:restricted-products}
As discussed above, the preceding results should be interpreted as statements about thermalization at infinite temperature. To get ensembles other than infinite temperature we must restrict the set of initial states we average over. One natural way is to still consider a partition into regions $R_s, s=1,2,...,S$, but in each region we restrict the state into a subspace of Hilbert space $\HS_s$, denoted as $\HS_{Ms}\subset \HS_s$. Physically, $\HS_{Ms}$ is the subspace of states in an energy window $E_0<E<E_0+\Delta E$, when we define the energy with respect to the subsystem Hamiltonian of $R_s$, neglecting the boundary term contribution. We can define a ``microcanonical'' density matrix $\rho_{Ms}=\pi_{Ms}/D_{Ms}$ for each region, with $\pi_{Ms}$ the projection operator onto $\HS_{Ms}$, and $D_{Ms}$ the dimension of $\HS_{Ms}$.  Then we consider the initial state as pure states drawn from the ensemble $\rho_M=\otimes_s\rho_{Ms}$, which are states with zero entanglement entropy between different regions, and have a finite energy density. It will be convenient to change the normalization on the operator inner product to be \(\langle A, B \rangle = \Tr[A^{\dagger} B]\).

The above results suggest that entropy growth and thermalization of these product states should be related to entanglement properties of some state depending on $U(t)$; tentatively, call this state $\ket{U_M(t)}$ (in Sections~\ref{sec:quantum_chaos} and \ref{sec:random-product}, the relevant state was isomorphic to the operator $U(t)$). As a first check that we have chosen a useful state, it is natural to require that some analog of \eqref{eq:i-rhou-bound} hold for $\rho^{U_M(t)}$. Such a result would suggest that correlations in $\ket{U_M(t)}$ are related in the same intuitive way to thermalization of states from the ensemble $\rho_M$ as correlations in the state $\ket{U(t)}$ are to thermalization of states from the ensemble $\matid/D$. A useful choice turns out to be \(\ket{U_M(t)} =\matid\otimes U(t) \rho_M^{1/2}\ket{I}\); since we have chosen $\rho_M$ compatible with the tensor factorization of $\HS$, to understand this state one can just put projectors on each input leg of $U(t)$ in Figures~\ref{fig:explicit-rho-u-construction} and \ref{fig:setup-formula}. The case \(\pi_M = \matid\) has been described in Section \ref{sec:quantum_chaos}; it turns out this is a very special case, due to the fact that \(\matid\) commutes with everything. For general \(\rho_M\) and two regions $A_F, B_P$ in the past and future systems, respectively, we have the bound
\begin{equation}
  \label{eq:i-um-corr-bound}
  I[\rho^{U_M(t)}; A_F, B_P] \ge \frac{1}{2} \left( 
    \frac{\left( \expv{O_A(t) O_B(0)}_{\rho_M} - \expv{O_A(t)}_{\rho_M} \expv{O_B(0)}_{\rho_M} \right) + \expv{O_A(t) [ \pi_M, O_B(0) ]}_{\rho_M}}{\|O_A\|\|O_B\|}
  \right)^2.
\end{equation}
If the commutator \(\left[\pi_M, O_B\right]\) is small, \eqref{eq:i-um-corr-bound} becomes exactly analogous to \eqref{eq:i-rhou-bound}. For example, suppose \(H\) has the form \(H \equiv H_L = \sum_s H_s + \sum_{\partial s} H_{\partial s}\) where \(H_s\) act on disjoint subsystems \(\HS_s\), and the boundary terms \(H_{\partial s}\) are allowed to couple ``nearby'' subsystems. If we then choose \(\pi_{Ms}\) to project onto some subsystem energy window (one where the eigenvalues of \(H_s\) lie in some fixed range) and take \(\pi_M = \bigotimes_s \pi_{Ms}\), operators \(O_B\) that are local to subsystems and do not change the energy outside the energy window will have zero commutator with \(\pi_M\). Another example is some local conserved quantity that we choose to concentrate in some subsystem \(\HS_s\) by choice of \(\pi_{Ms}\); if \(O_B\) does not transport this charge across subsystems it will have zero commutator with \(\pi_M\). If the above conditions are only met approximately (\(O_B\) has small matrix elements for bringing states out of and into \(\HS_M\)), the commutator will be small. Of course, we can enforce a zero-commutator condition on \(O_B\) by simply taking it to \(\tilde{O}_B = \pi_M O_B \pi_M\). Then we can directly interpret \eqref{eq:i-um-corr-bound} as the ``re-equilibration'' of \(\rho_M\) after acting by \(O_B\); perturbing the state by \(O_B\) does not affect the action of \(O_A\) in the future. Of course, the case \(\pi_M = \matid\) reduces exactly to \eqref{eq:i-rhou-bound} for any choice of \(O_B\). 

As mentioned, \(\pi_M = \matid\) is a very special case, and this gives rise to important differences when relating information measures to chaos and equilibration for generic \(\rho_M\). The bound \(S^{(2)} \le S\) is always true, so \(S^{(2)}\) of \(\rho^{U_M(t)}\) is still a good measure of the ``correlation'' between the past and future. Important special cases are \(S^{(2)}[\rho^{U_M(t)}_{A_F}] = S^{(2)}[\rho_{MA}(t)]\) and \(S^{(2)}[\rho^{U_M(t)}_{B_P}] = S^{(2)}[\rho_{MB}(0)]\). On the other hand, since \(\rho_M\) can have non-trivial time evolution, the quantity \(S^{(2)}[\rho^{U_M(t)}_{A_F}] + S^{(2)}[\rho^{U_M(t)}_{B_P}] - S^{(2)}[\rho^{U_M(t)}_{A_F \cup B_P}]\) can become negative (in contrast to the \(\pi_M = \matid\) case), so \(I^{(2)}\) as defined in Section \ref{sec:quantum_chaos} is not as fundamental a quantity. It also does not bound the corresponding mutual information \(I\). We \emph{can} define a quantity that upper bounds \(I\), 
\begin{eqnarray}
\tilde{I}[\rho^{U_M(t)}; A_F, B_P] &=&I[\rho^{U_M(t)}; A_F, B_P] + (S[\rho^{U_M(t)}_{A_F \cup B_P}] - S^{(2)}[\rho^{U_M(t)}_{A_F \cup B_P}])\nonumber\\
&=&S[\rho^{U_M(t)}; A_F]+S[\rho^{U_M(t)}; B_P]-S^{(2)}[\rho^{U_M(t)}_{A_F \cup B_P}]
\end{eqnarray}

For \(\pi_M = \matid\), \(\tilde{I}[\rho^{U_M(t)}; A_F, B_P] = I^{(2)}[\rho^{U_M(t)}; A_F, B_P]\). There are equalities analogous to \eqref{eq:unitary-renyi-otoc} and \eqref{eq:unitary-renyi-2pt} relating \(\tilde{I}\) to chaos:
\begin{align}
  \label{eq:finite-temp-i-chaos}
  e^{\tilde{I}[\rho^{U_M(t)}; A_F, R_P]} 
  &= \left( \frac{e^{S[\rho_{MA}(t)]}}{D_A} \right) \left( \frac{D_{\bar{R}}}{D_{M \bar{R}}} \right)
    D_A^2 C_4(\mathcal{O}_A(t), \tilde{\mathcal{O}}_{\overline{R}}(0))_{\rho_M} \\
  \label{eq:finite-temp-i-2pt}
  &= \left( \frac{e^{S[\rho_{MA}(t)]}}{D_A} \right) \left( \frac{D_{MR}}{D_R} \right) (D_A D_R \expv{\mathcal{O}_A(t) \mathcal{O}_R(0)}_{\rho_M})^2,
\end{align}
for regions \(R \in \mathcal{P}(\{R_s\})\) (for other sorts of regions, factors of \(D_{MR}\) will be replaced by entropies), where \(\tilde{\mathcal{O}}_R = \pi_M \mathcal{O}_R \pi_M\). This modification of \(\mathcal{O}_R\) has a natural interpretation, paralleling the discussion of \eqref{eq:i-um-corr-bound}. The OTOCs in \eqref{eq:finite-temp-i-chaos} are to be computed for operators in the past that do not move states out of \(\HS_M\) (and act by zero on states outside); in the examples following \eqref{eq:i-um-corr-bound}, \(\tilde{\mathcal{O}}_R\) will conserve local energy density or subsystem charge, respectively. We also clearly have \([\pi_M, \tilde{\mathcal{O}}_R] = 0\), so according to \eqref{eq:i-um-corr-bound} the mutual information bounds the effect of these operators in the most intuitive way. Note that as \(\HS_{Ms}\) becomes smaller, the four-point contributions \eqref{eq:finite-temp-i-chaos} become more important.

With these preparations, we can extend the results of Section \ref{sec:random-product} to the case of generic \(\rho_M\). For product states \(\ket{\psi_M} = \bigotimes_s \ket{a_{Ms}}\) with each \(\ket{a_{Ms}}\) taken from \(\HS_{Ms}\), we obtain the result analogous to \eqref{eq:infinite-temp-renyi2}:
\begin{multline}
  \label{eq:finite-temp-renyi2}
  \haarexp[\exp (-S^{(2)}[\rho^{\psi_M}_A(t)])]
  = \frac{1}{\prod_s (1 + 1 / D_{Ms})} \\
  e^{- S^{(2)}[\rho_{MA}(t)]} \left( 1 + \sum_{\substack{R \in \mathcal{P}(\{R_s\}) \\ R \ne \varnothing}} e^{-(S[\rho_{MA}(t)] - S^{(2)}[\rho_{MA}(t)])} \frac{e^{\tilde{I}[\rho^{U_M(t)}; A_F, R_P]}}{D_{MR}} \right).
\end{multline}
The \(\tilde{I}\) terms are the positive corrections for product states from the entropy computed from \(\rho_M\). We can also generalize the discussion surrounding \eqref{eq:one-norm-inft-t} to see that
\begin{equation}
  \label{eq:finite-temp-one-norm}
  \haarexp \left[ \| \rho^{\psi_M}_A(t) - \rho_{MA}(t) \|_1 \right]
  \le \sqrt{D_A \left( \haarexp e^{- S^{(2)}[\rho^{\psi_M}_A(t)]} - e^{- S^{(2)}[\rho_{MA}(t)]}\right)},
\end{equation}
so states chosen from the ensemble given by \(\rho_M\) ``equilibriate'' (in the above sense of 1-norm) to \(\rho_M\) given small \(\tilde{I}\) terms. The major difference is that \(\rho_M\) is generically not thermal. Interpreting small \(\tilde{I}\) as chaos, this shows that chaos is sufficient to ``scramble'' initial conditions to the extent that the particular state within the initial ensemble is irrelevant, but we have not shown that the ``unentangled microcanonical ensemble'' \(\rho_M\) itself thermalizes. That said, in a system with a local Hamiltonian and with a choice of the regions $R_s$ of size much bigger than the thermal correlation length, the contribution of boundary terms to energy is small, and $\rho_M$ has a volume law entropy that is close to the thermal value at the same energy expectation value. In other words, an initial pure state drawn from $\rho_M$ has already almost thermalized when the reduced density matrix approaches that of $\rho_M$.

\section{Conclusion}
\label{sec:conclusion}
We have explored the consequences of small correlation (as computed in \(U(t)\) and \(U_M(t)\)) between the past and future. For the density matrices \(\rho_{\beta = 0}\) and \(\rho_M\) the expressions \eqref{eq:i-rhou-bound} and \eqref{eq:i-um-corr-bound} respectively show that certain types of perturbations to these density matrices in some region \(B\) are ``forgotten'' in some region \(A\) as long as the information between \(A\) in the past and \(B\) in the future for \(U(t)\) or \(U_M(t)\) has had time to decay. Next, \eqref{eq:infinite-temp-renyi2} and \eqref{eq:finite-temp-renyi2} show that the decay of information between past and future regions corresponds to entropy growth for far-from-equilibrium pure states. Finally, using these expressions in combination with \eqref{eq:one-norm-inft-t}, \eqref{eq:fluct-bound}, and \eqref{eq:finite-temp-one-norm}, we have shown that this decay of information between past and future means the initial conditions of a particular pure state chosen from an ensemble are forgotten (although the ensemble itself is not). Although the discussion proceeds most naturally in terms of information, we can also conclude that quantum chaos as diagnosed by the OTOC and the decay of local two-point functions imply entropy growth and erasure of initial conditions (``equilibration'') by relating the OTOC and two-point functions to information. Likewise, generic thermalization implies the contribution of \(I^{(2)}\) or \(\tilde{I}\) terms in \eqref{eq:infinite-temp-renyi2} or \eqref{eq:finite-temp-renyi2} are small, so mutual information between local regions in the future and sub-extensive regions in the past is bounded above. Thus there is a sense in which quantum thermalization implies chaos.

The most important extension of this work is a deeper understanding of the finite temperature results. The same state may be a member of several ensembles \(\rho_M\), but the formalism we developed does not identify a preferred density matrix. Such a preferred density matrix should be a time independent distribution, for example the Boltzmann distribution, when that pure state equilibriates. A first step may be to find conditions such that that the ``more thermal'' (higher entanglement) \(\rho_M\) thermalize. It should also be possible to improve the factor \(D_A\) in \eqref{eq:finite-temp-one-norm} in the case that \(\haarexp e^{- S^{(2)}[\rho^{\psi_M}_A(t)]} \simeq e^{- S^{(2)}[\rho_{MA}(t)]}\) for locally thermalizing \(\rho_M\). Finally, to make this work more practically applicable, it is important to show that either chaos as measured by the OTOC or decay of information between the future and past is generic for local Hamiltonians.
\section{Acknowledgements}
\label{sec:acknowledgements}
We would like to thank Michael Walter and Beni Yoshida for useful discussions. This work is supported by the National Science Foundation through the grant No. PHY-1720504 (YL and XLQ), and by the Hertz Foundation (YL).

\printbibliography
\appendix

\section{Derivation of main results}
\label{sec:deriv-main-results}
We present the derivation of the results in Section~\ref{sec:restricted-products}, which are a strict generalization of the results in Sections~\ref{sec:quantum_chaos} and \ref{sec:random-product}. The main tool is the Schur-Weyl duality, which describes the combined action of the symmetric and unitary groups on tensor product spaces. On the vector space $V_n = (\C^D)^{\otimes^n}$, the symmetric group on $n$ letters, $\SymG^n$, acts in a natural way by permuting the $n$ factors, while the unitary group $U(D)$ acts by $U^{\otimes^n}$ for $U \in U(D)$.
\begin{theorem}[Schur-Weyl Duality]
  \label{thm:schur-weyl}
  Under the combined natural actions of $\SymG^n$ and $U(D)$ on the vector space $V_n$ (as defined above), $V_n$ can be written as a direct sum
  \begin{equation*}
    V_n = \bigoplus_Y W_Y \otimes S_Y
  \end{equation*}
  where $Y$ is an index running over Young diagrams with $n$ boxes, and $W_Y$ ($S_Y$) is an irreducible representation of $U(D)$ ($\SymG^n$) not isomorphic to any other representation appearing with different $Y$.
\end{theorem}
We will typically use this theorem to constrain operators that commute with the action of $U(D) \times \SymG^n$; since each irrep of the combined action appears only once in the decomposition of $V_n$, such an operator must act as multiplication by a constant on each irrep by Schur's lemma. Furthermore, the theorem tells us we can project onto each irrep by projecting onto an irrep of only $\SymG^n$, so each such operator can be written as a sum of projectors, each of which is in turn a sum of elements of $\SymG^n$.

As an intermediate result, we must compute the Haar integral $A_n = \int dU (U \ket{\psi} \bra{\psi} U^{\dagger})^{\otimes^n}$, which is clearly independent of the choice of $\ket{\psi}$. Furthermore, $A_n$ commutes with the above actions of $U(D)$ and $\SymG^n$, so we take $A_n$ to be a sum of $\sigma \in \SymG^n$. It is easy to check that in fact $\sigma A_n = A_n$ for all $\sigma \in \SymG^n$, so $A_n \propto \sum_{\sigma \in \SymG^n} \sigma$. To find the normalization factor, note that \[\Tr[\sum_{\sigma \in \SymG^n} \sigma] = \sum_{c = 1}^n (\text{count of permutations with } c \text{ cycles}) D^c = \prod_{m = 0}^{n - 1} (D + m)\]~\footnote{The last equality is proved easily after noting that for a given permutation of $n - 1$ elements, to form a permutation of $n$ elements the $n$th element either forms a new cycle, contributing a factor of $D$, or can be put in an existing cycle in $n - 1$ distinct ways (regardless of the permutation).}. This gives
\begin{equation*}
  \haarexp[\ket{\psi} \bra{\psi}] = \int dU (U \ket{\psi} \bra{\psi} U^{\dagger})^{\otimes^n}
  = \left[ \prod_{m = 1}^{n - 1} (1 + m / D) \right]^{-1} \frac{1}{D^n} \sum_{\sigma \in \SymG^n} \sigma.
\end{equation*}

We now compute the expectation of the second Renyi entropy in the setup of Section~\ref{sec:restricted-products}. Call the non-identity element of $\SymG^2$, that swaps tensor factors, $\swap$. If we have a Hilbert space $\HS$ with a subsystem labelled $A$, there is a permutation group $\SymG^n_A$ that acts on $\HS^{\otimes^n}$ by only permuting tensor factors corresponding to subsystem $A$ between copies. We refer to these group elements by a subscript $A$, so for example $\swap_A$. To compute the expectation of the second Renyi entropy, we use the relation $\Tr[\rho_A^2] = \Tr[\rho^{\otimes^2} \swap_A]$. As a reminder, the distribution on initial $\rho^{\psi_M}$ in our case is fixed as follows. We are given a partition of Hilbert space into $S$ subsystems, $\HS = \bigotimes_s \HS_s$, and in the vector space associated to each subsystem we choose a linear subspace $\HS_{Ms}$ (with associated projector $\pi_{Ms}$). The distribution on $\rho^{\psi_M}$ is independently Haar random on the subspace of each subsystem. From the above discussion on the Haar integral, it follows that
\begin{equation*}
  \haarexp[\rho^{\psi_M}(0)^{\otimes^2}] = \frac{1}{\prod_s (1 + 1/D_{Ms})} \rho_M^{\otimes^2} \prod_s (\matid_{s} + \swap_s)
  = \frac{1}{\prod_s (1 + 1/D_{Ms})} \rho_M^{\otimes^2} \sum_{R \in \mathcal{P}(\{R_s\})} \swap_R,
\end{equation*}
where $D_{Ms}$ is the dimension of $\HS_{Ms}$, $\rho_M = \prod_s \pi_{Ms} / D_{Ms}$, and $\mathcal{P}(\{R_s\})$ is the powerset of subsystems. The second equality comes from noting that the product has $2^S$ terms, based on a choice of $\matid_s$ or $\swap_s$ for each subsystem, and the included swaps combine to give a single swap of all included subsystems. We then have
\begin{align}
  \label{eq:s2-haar-exp-main-deriv}
  \haarexp\left[ \Tr[\rho^{\psi_M}(t)_A^2] \right]
  &= \Tr[U(t)^{\otimes^2} \haarexp[\rho^{\psi_M}(0)^{\otimes^2}] (U(t)^{\dagger})^{\otimes^2} \swap_A] \\
  \label{eq:s2-swap-expression}
  &= \frac{1}{\prod_s (1 + 1/D_{Ms})} \sum_{R \in \mathcal{P}(\{R_s\})} \Tr[(U(t) \sqrt{\rho_M})^{\otimes^2} \swap_R (U(t) \sqrt{\rho_M})^{\dagger \otimes^2} \swap_A] \\
  \label{eq:s2-haar-exp-as-s2-deriv}
  &= \frac{1}{\prod_s (1 + 1/D_{Ms})} \sum_{R \in \mathcal{P}(\{R_s\})} e^{-S^{(2)}[\rho^{U_M(t)}_{A_F \cup R_P}]}
\end{align}
where $U_M(t) = U(t) \sqrt{\rho_M}$. To see the last equality, we refer to the explicit construction of the state corresponding to $U_M(t)$ following the procedure of Figure~\ref{fig:explicit-rho-u-construction} to check that the index contractions are correct, and the proportionality factor is correct since $\Tr[U(t) \rho_M U(t)^{\dagger}] = 1$. By the same construction, we can compute $S^{(2)}[\rho^{U_M(t)}_{R_P}] = S[\rho^{U_M(t)}_{R_P}] = \ln D_{MR}$, $S^{(2)}[\rho^{U_M(t)}_{A_F}] = S^{(2)}[\rho_{MA}(t)]$, and $S[\rho^{U_M(t)}_{A_F}] = S[\rho_{MA}(t)]$, so that upon multiplication by appropriate factors of entropy, equations \eqref{eq:infinite-temp-renyi2} and \eqref{eq:finite-temp-renyi2} follow from \eqref{eq:s2-haar-exp-as-s2-deriv}.

To connect \eqref{eq:s2-haar-exp-main-deriv}, and more generally entropies of $\rho^{U_M(t)}$, to observables, we use the explicit form of projectors onto irreps of $\SymG^2$: $\pi_{\pm} = (\matid \pm \swap) / 2$. Then an operator $A$ on $\HS^{\otimes^2}$ that commutes with the joint action of $U(D) \times \SymG^2$ can explicitly be written as a sum of $\pi_{\pm}$, with coefficients $\Tr[A \pi_{\pm}] / \Tr[\pi_{\pm}]$. This gives
\begin{equation*}
  A = \frac{1}{D^2 - 1} \left( (\Tr[A] - \Tr[A \swap] / D) \matid + (\Tr[A \swap] - \Tr[A] / D) \swap \right).
\end{equation*}
In particular, we can implement $\swap$ in terms of any operators that commute with $U(D) \times \SymG^2$ and have $D\Tr[A] = \Tr[A \swap]$. One example can be found by taking a complete basis of Hermitian operators orthonormal under the inner product $\langle A, B \rangle = \Tr[A^{\dagger} B] / D$, $\mathcal{B}$ (we make this normalization choice so that eigenvalues of operators in the basis can be $\pm 1$). Since conjugation by a unitary $U \in U(D)$ preserves Hermeticity and the inner product, conjugating every element of $\mathcal{B}$ gives a new, still orthonormal basis of the real vector space of Hermitian operators, so can be written as a real orthogonal matrix acting on the elements of $\mathcal{B}$. Thus the average over Hermitian operators $Y = D^{-2} \sum_{\mathcal{O} \in \mathcal{B}} \mathcal{O}^{\otimes^2}$ is invariant under conjugation by a unitary (and is therefore independent of the particular choice of operator basis), and clearly commutes with the action of $\SymG^2$; in other words $Y$ commutes with the joint action of $U(D) \times \SymG^2$. Taking a basis including the identity shows $\Tr[Y] = 1$, and the normalization condition gives $\Tr[Y \swap] = D^{-2} \sum \Tr[\mathcal{O}^2] = D$, in other words $Y = \swap / D$. We can then write (using \eqref{eq:s2-swap-expression} and the following discussion), assuming that $R$ factors through the tensor factorization into $\HS_s$ for convenience,
\begin{align*}
  e^{-S^{(2)}[\rho^{U_M(t)}_{A_F \cup R_P}]}
  &= \Tr[U(t)^{\dagger \otimes^2} \swap_A U(t)^{\otimes^2} \swap_R \rho_M^{\otimes^2}]
  = D_A D_R \Tr[(\mathcal{O}_A(t) \mathcal{O}_R \rho_M)^{\otimes^2}] = D_A D_R \left( \langle \mathcal{O}_A(t) \mathcal{O}_R \rangle_{\rho_M} \right)^2 \\
  &= D_A \Tr[\mathcal{O}_A(t)^{\otimes^2}\swap_{\overline{R}} \rho_M^{\otimes^2} \swap] \\
  &= \frac{D_A D_{\overline{R}}}{D_M} \Tr[\mathcal{O}_A(t) \mathcal{O}_{\overline{R}} \pi_M \mathcal{O}_A(t) \mathcal{O}_{\overline{R}} \rho_M] \\
  &= \frac{D_A D_{\overline{R}}}{D_M} \Tr[\mathcal{O}_A(t) \tilde{\mathcal{O}}_{\overline{R}} \mathcal{O}_A(t) \tilde{\mathcal{O}}_{\overline{R}} \rho_M]
  = \frac{D_A D_{\overline{R}}}{D_M} C_4(\mathcal{O}_A(t), \tilde{\mathcal{O}}_{\overline{R}}(0))_{\rho_M}
\end{align*}
where the first equality on the last line follows since $[\pi_M, \swap_R] = 0$. These expressions, after multiplication by $D_{MR} e^{S[\rho_{MA}(t)]}$, give equations \eqref{eq:unitary-renyi-otoc},\eqref{eq:unitary-renyi-2pt},\eqref{eq:finite-temp-i-chaos}, and \eqref{eq:finite-temp-i-2pt}. Finally, equations \eqref{eq:i-rhou-bound} and \eqref{eq:i-um-corr-bound} are direct consequences of the well-known inequality
\begin{equation*}
  I[\rho; A, B] \ge \frac{1}{2} \left( \frac{\expv{O_A O_B}_{\rho} - \expv{O_A}_{\rho} \expv{O_B}_{\rho}}{\| O_A \| \| O_B \|} \right)^2
\end{equation*}
and the definition of $\rho^{U_M(t)}$.

As a possible tool for computing higher Renyi entropies, we note that elements of $\SymG^n$ can be implemented in terms of averages of local operators for $n > 2$, despite the fact that we used a property special to $n = 2$ (invariance of $\sum \mathcal{O}^{\otimes^2}$ when summed over an orthonormal basis) above. The idea is to take some as-yet unchosen Hermitian operator $A$, and define $M = \int dU (U A U^{\dagger})^{\otimes^n}$, which commutes with $U(D) \times \SymG^n$, and so is a weighted sum of projectors onto irreps of $\SymG^n$. The numbers $\Tr[M \sigma]$ for $\sigma \in \SymG^n$ determine the weights; these are in turn products of traces of powers of $A$. Thus $M$ depends only on the spectrum of $A$, and by tuning this spectrum we can tune the weights of projectors. For example, if we take $n = 2$, the Haar average over operators $A$ with some fixed spectrum satisfying $D (\sum_i \lambda_i)^2 = \sum_i \lambda_i^2$ is proportional to $\swap$.

\section{Derivation of Equation~\eqref{eq:i-bound-i2}}
\label{sec:deriv-mi-bound}
We present a short derivation of \eqref{eq:i-bound-i2}, based on equation 23 of \autocite{harremoes2001inequalities}. That equation, in our notation, is
\begin{equation*}
  S[\rho^{U(t)}_{A_F \cup B_P}] \le \ln (D_A D_B) \left( 1 - \tau \left( \frac{e^{- S^{(2)}[\rho^{U(t)}_{A_F \cup B_P}]} - (D_A D_B)^{-1}}{1 - (D_A D_B)^{-1}} \right) \right),
\end{equation*}
where another result of \autocite{harremoes2001inequalities} is that $\tau \ge \ln (D_A D_B) / (\ln (D_A D_B) + 1)$. Then, noting that $S[\rho^{U(t)}_{A_F}] = S^{(2)}[\rho^{U(t)}_{A_F}] = \ln D_A$ and likewise for $\rho^{U(t)}_{B_F}$ as $\rho^{U(t)}$ is maximally entangled between past and future (see Figure~\ref{fig:rho-u}), we can multiply both sides by $D_A D_B$ and use the bound on $\tau$ to obtain \eqref{eq:i-bound-i2}.

\end{document}